# Electron Spectra for Twisted Electron Collisions

A. Plumadore and A. L. Harris*

Physics Department, Illinois State University, Normal, IL, USA 61790

## Abstract

Ionization collisions have important consequences in many physical phenomena, and the mechanism that leads to ionization is not universal. Double differential cross sections (DDCSs) are often used to identify ionization mechanisms because they exhibit features that distinguish close collisions from grazing collisions. In the angular DDCS, a sharp peak indicates ionization through a close binary collision, while a broad angular distribution points to a grazing collision. In the DDCS energy spectrum, electrons ejected through a binary encounter collision result in peak at an energy predicted from momentum conservation. These insights into ionization processes are well-established for plane wave projectiles. However, the recent development of sculpted particle wave packets reopens the question of how ionization occurs for these new particle wave forms. We present theoretical DDCSs for (e,2e) ionization of atomic hydrogen for electron vortex projectiles. Our results predict that the ionization mechanism for vortex projectiles is similar to that of non-vortex projectiles, but that features in the DDCS that distinguish ionization mechanisms are obscured by the projectile's momentum uncertainty. Additionally, the projectile's non-zero transverse momentum increases the cross section for high energy ejected electrons.

## 1. Introduction

Ionization collisions provide valuable information about atomic structure, as well as charged particle dynamics. The measurement and theoretical prediction of electron spectra can indicate mechanisms that lead to target break up. In particular, double differential cross sections (DDCSs) for electron-impact ionization show distinct features that allow for the separation of soft collisions and binary encounter collisions. The angular distribution of DDCSs (ADDCSs) for soft collisions is broad and flat, while for binary encounter collisions it is narrow and sharply peaked [1–5]. Typically, the shape of the ADDCS is correlated with ejected electron energy, such that lower energy ejected electrons result in broader angular distributions due to the influence of the target electron's momentum uncertainty.

* alharri@ilstu.edu

The energy distribution of DDCSs (EDDCSs) can show signatures of the Bethe binary encounter peak and have been used to test the range of validity of the Born approximation and the semiclassical binary encounter approximation [1]. Along the bound electron Bethe ridge, the bound electron is at rest and is ionized through a purely binary collision [3,5]. In this case, the energy and momentum transfer of the projectile is completely absorbed by the bound electron.

The degree to which the Born approximation or binary encounter approximation is able to accurately predict experimental measurements indicates whether ionization occurs predominantly through binary collisions or other mechanisms. Several decades of experimental and theoretical investigations have shown that the binary encounter approximation is good near the peak of the ADDCS [1]. For large ejected electron energy, strong peaks in the experimental ADDCSs provide evidence that very little momentum is transferred to the nucleus [6] and indicate that the Born approximation and binary encounter approximation are valid for these kinematics. Further evidence of the validity of the Born approximation for fast ejected electrons is found in the good agreement between experiment and theory for both helium and molecular hydrogen targets, including the prediction of the binary encounter peak in the EDDCS [3].

To date, all of the studies of energy or angular electron spectra were completed using traditional, plane-wave type electrons in which the dynamics of (e,2e) collisions for simple atomic targets are now essentially well-understood [7–10]. However, in the last decade the production of new structured electron wave packets has led to a renewed interest in (e,2e) collisions for simple atomic targets [11–25]. In particular, electron vortex projectiles have many properties that make them quite different from their plane wave counterparts. For example, they have non-zero transverse linear momentum and can carry quantized angular momentum. These unique properties have been proposed to have uses in nanoscience [26–29], microscopy [17,26,30], and the study of fundamental atomic properties [26,27,31]. Key to realizing these applications is a thorough understanding of how electron vortex projectiles interact with matter. Collision physics, with its long history of providing key insights into atomic and molecular structure and few body interactions, can provide some of the necessary fundamental information required to understand the interactions of sculped electrons with matter.

To date, studies using electron vortex projectiles have been exclusively theoretical and include numerous predictions that can be experimentally tested. For example, in [21] a study of the excitation of hydrogen by electron vortex projectiles showed that orbital angular momentum is transferred directly from the projectile to the atom. In [24], it was shown that chirality dependent scattering can be generated and that an optimized choice of external magnetic field and vortex topological charge can enhance the observed dichroism effects. In one of our previous studies, we showed that ionization into the perpendicular plane is significantly enhanced with vortex projectiles [32]. Additionally, in [11–13], we showed that the triple differential cross sections for electron vortex ionization of ground and excited state hydrogen targets were qualitatively and quantitatively different than those of non-vortex projectiles. Here, we expand upon our prior work by exploring grazing and close collisions for twisted electrons.

In this work, we present angular and energy distributions of DDCSs for (e,2e) collisions between electron vortex projectiles and hydrogen atoms. Because DDCSs are more feasible for experimentalists, while still yielding valuable information about the scattering mechanisms of twisted electrons, our results will hopefully spur experimental interest in twisted electron collisions. Our calculations show that the ionization mechanism for vortex projectiles is similar to that of non-vortex projectiles, but that the projectile's momentum uncertainty causes a broadening and splitting of the ADDCS peak. Additionally, the projectile's non-zero transverse momentum increases the cross section for high energy ejected electrons. Atomic units are used throughout.

## 2. Theory

### A. Triple and Double Differential Cross Sections

In a kinematically complete collision, the momenta of all free particles are known. Theoretically, this requires calculation of the triple differential cross section (TDCS), which is differential in the solid angles of both final state electrons, as well as the ionized electron's energy

$$TDCS = \frac{d^3\sigma}{d\Omega_1 d\Omega_2 dE_2}. \tag{1}$$

Experimentally, a measurement of the TDCS requires two of the three final state particles to be detected in coincidence.

For double differential cross sections, only the momentum of one of the final state electrons is known and the theoretical TDCS must be integrated over the solid angle of the undetected electron

$$DDCS = \int TDCS \, d\Omega_1 = \frac{d^2\sigma}{d\Omega_2 dE_2}. \tag{2}$$

Experimentally, the DDCS requires the detection of only one of the final state particles. Double differential cross sections are therefore differential in electron and solid angle of one of the ejected electrons.

Electron vortex projectiles are experimentally generated in electron microscopes at energies on the order of a few to hundreds of keV. For such fast projectiles, the first Born Approximation (FBA) is appropriate for the calculation of the cross sections. Within the FBA, the

TDCS can be written in terms of direct $f$ and exchange $g$ amplitudes that account for the indistinguishability of the two electrons in the final state

$$\frac{d^3\sigma}{d\Omega_1 d\Omega_2 dE_2} = \mu_{pa}^2 \mu_{ie} \frac{k_f k_e}{k_i} \frac{1}{2} (|f^V|^2 + |g^V|^2 + |f^V - g^V|^2). \quad (3)$$

where $\mu_{pa}$ and $\mu_{ie}$ are the reduced masses of the projectile and target atom and the proton and ionized electron respectively. The incident and scattered projectile momenta are $\vec{k}_i$ and $\vec{k}_f$, and the ejected electron momentum is $\vec{k}_e$.

### B. Transition Matrices

The direct and exchange vortex transition matrices are given by

$$f^V = N^V \langle \chi_{\vec{k}_f}(\vec{r}_1) \chi_{\vec{k}_e}(\vec{r}_2) | V_i | \chi_{\vec{k}_i}(\vec{r}_1) \Phi(\vec{r}_2) \rangle \quad (4)$$

and

$$g^V = N^V \langle \chi_{\vec{k}_f}(\vec{r}_2) \chi_{\vec{k}_e}(\vec{r}_1) | V_i | \chi_{\vec{k}_i}(\vec{r}_1) \Phi(\vec{r}_2) \rangle \quad (5)$$

where $\chi_{\vec{k}_i}(\vec{r}_1)$ is the incident projectile vortex Bessel wave function and $\Phi(\vec{r}_2)$ is the target atom wave function, which is known analytically. The vectors $\vec{r}_1$ and $\vec{r}_2$ are the lab frame position vectors for the projectile and atomic electron, respectively. In the final state, both outgoing electrons are assumed to be non-vortex electrons. For the scattered projectile $\chi_{\vec{k}_f}(\vec{r}_{1,2})$, we use a plane wave, and for the ionized electron wave function $\chi_{\vec{k}_e}(\vec{r}_{1,2})$, we use either a plane wave or Coulomb wave. The normalization factor is given by $N^V = -(2\pi)^{3/2}$ and the perturbation is

$$V_i = -\frac{1}{r_1} + \frac{1}{r_{12}}, \quad (6)$$

which is the Coulomb interaction between the projectile and target atom.

The collision system used to calculate the transition matrices of Eqs. (4) and (5) consists of an incident projectile with momentum $\vec{k}_i$ that is propagating along the z direction. Following the collision, the projectile has a momentum $\vec{k}_f$ and is scattered at a polar angle $\theta_1$ and azimuthal angle $\varphi_1$. The ionized electron is emitted with energy (momentum) $E_2$ ($\vec{k}_e$) with polar coordinate $\theta_e$ and azimuthal coordinate $\varphi_e$.

The delta function normalized plane wave is given by

$$\chi_{\vec{k}}(\vec{r}) = \frac{e^{i\vec{k}\cdot\vec{r}}}{(2\pi)^{3/2}}, \tag{7}$$

and the Coulomb wave is given by

$$\chi_{\vec{k}_e}(\vec{r}_2) = \Gamma(1-i\eta)e^{-\frac{\pi\eta}{2}}\frac{e^{i\vec{k}_e\cdot\vec{r}_2}}{(2\pi)^{3/2}}F_1\big(i\eta, 1, -ik_e r_2 - i\vec{k}_e\cdot\vec{r}_2\big), \tag{8}$$

where $\Gamma(1-i\eta)$ is the gamma function and $\eta$ is the Sommerfeld parameter.

Like the plane wave, the field-free incident Bessel electron wave function used for the vortex projectile is also a solution to the free particle Schrödinger equation and is given by

$$\chi_{\vec{k}_i}(\vec{r}) = \frac{e^{il\varphi}}{2\pi}J_l(k_{i\perp}\rho)e^{ik_{iz}z}, \tag{9}$$

where the electron's spatial coordinates are expressed in cylindrical coordinates $(\rho, \varphi, z)$ and $l$ is the electron's quantized orbital angular momentum, also known as the topological charge. Unlike the plane wave, the Bessel wave function for an incident electron propagating along the z-axis has both a longitudinal $k_{iz}$ and transverse $k_{i\perp}$ momentum. These momenta can be written in terms of the beam's opening angle $\alpha$ as

$$k_{i\perp} = k_i \sin\alpha, \tag{10}$$

and

$$k_{iz} = k_i \cos\alpha. \tag{11}$$

Equation (9) shows that the Bessel wave function explicitly depends on the projectile transverse momentum. However, only the magnitude of the incident momentum is specified, not its azimuthal angle. Therefore, there is an inherent uncertainty in the incident projectile momentum, which can cause uncertainty in the ionized electron momentum [11–13,32].

### C. Average over impact parameters

Unlike their plane wave counterparts, vortex projectiles are non-uniform in the transverse direction. In particular, the Bessel wave functions used in this work have a phase singularity at a given spatial location that results from their helical wave fronts [33–35]. Their helicity also leads to their orbital angular momentum, which for non-zero values results in a spatial node along the propagation axis [36]. The spatial node and non-uniformity in the transverse direction lead to the need to define an impact parameter $\vec{b}$ to describe the transverse location of the phase singularity relative to the z-axis.

A useful expression for the Bessel wave function with arbitrary impact parameter is to write it as a superposition of tilted plane waves [21]

$$\chi_{\vec{k}_i}(\vec{r}_1) = \frac{(-i)^l}{(2\pi)^2} \int_0^{2\pi} d\phi_{ki}\, e^{il\phi_{ki}} e^{i\vec{k}_i \cdot (\vec{r}_1 - \vec{b})}, \tag{12}$$

where $\phi_{ki}$ is the azimuthal angle of the incident projectile momentum. Substitution of this expansion into the direct and exchange transition matrices of Eqs. (4) and (5) yields

$$f^V(\vec{q}, \vec{b}) = \frac{(-i)^l}{(2\pi)} \int_0^{2\pi} d\phi_{ki}\, e^{il\phi_{ki}} f^{PW}(\vec{q}) e^{-i\vec{k}_i \cdot \vec{b}} \tag{13}$$

and

$$g^V(\vec{q}, \vec{b}) = \frac{(-i)^l}{(2\pi)} \int_0^{2\pi} d\phi_{ki}\, e^{il\phi_{ki}} g^{PW}(\vec{s}) e^{-i\vec{k}_i \cdot \vec{b}}, \tag{14}$$

where $f^{PW}(\vec{q})$ and $g^{PW}(\vec{s})$ are the non-vortex direct and exchange amplitudes, $\vec{q} = \vec{k}_i - \vec{k}_f$ is the direct momentum transfer vector and $\vec{s} = \vec{k}_i - \vec{k}_e$. These two expressions show that the vortex transition matrices can be written as a superposition of the non-vortex transition matrices.

For traditional collision experiments, the impact parameter for a given collision event cannot be determined, and therefore theory must average over all impact parameters in order to compare with experiment. Using Eqs. (13) and (14) and averaging the TDCS over impact parameter yields

$$\frac{d^3\sigma}{d\Omega_1 d\Omega_2 dE_2} = \mu_{pa}^2 \mu_{pi} \frac{k_f k_e}{(2\pi)k_{i_z}} \int d\phi_{k_i} (|f^{PW}(\vec{q})|^2 + |g^{PW}(\vec{s})|^2 + |f^{PW}(\vec{q}) - g^{PW}(\vec{s})|^2), \quad (15)$$

in which the vortex TDCS is written as a superposition of the non-vortex TDCS. Likewise, the vortex DDCS can also be written as a superposition of the non-vortex DDCS

$$DDCS^V = \int d\phi_{k_i} DDCS^{NV}. \quad (16)$$

For the impact parameter averaged cross sections, there is no dependence on the vortex projectile's topological charge, and therefore any orbital angular momentum information is washed out [11–13,22]. The cross section's dependence on the projectile's vortex properties is thus limited to its momentum and opening angle.

### 3. Results

For a fixed incident projectile energy, the DDCS depends on one of the final state electron's energy and solid angle. Here, we present both the DDCS as a function of energy for fixed solid angle (EDDCS) and the DDCS as a function of solid angle for fixed energy (ADDCS), as they provide different information regarding the scattering mechanism. We have chosen an incident projectile energy of 1 keV, which is sufficiently high to ensure validity of the FBA. With the exception of $E_2 = E_f = 493.2 \, eV$, the energy sharing is highly asymmetric, such that the exchange amplitude can be safely neglected for the ADDCS, however it must be included for the EDDCS because $E_2$ becomes larger than $E_f$. For the ADDCS, the use of a Coulomb wave to describe the slow ionized electron is important, however a plane wave treatment is sufficient for the EDDCS. A more detailed comparison of the different FBA models and explanation of our choice of approximations can be found in Appendices A and B.

## A. Angular Double Differential Cross Sections

From classical momentum conservation for a binary collision between a non-vortex projectile and target electron, the ejected electron's momentum can be written as

$$\vec{k}_e = \vec{q} + \vec{k}_{eb} \tag{17}$$

where $\vec{q} = \vec{k}_i - \vec{k}_f$ is the projectile momentum transfer and $\vec{k}_{eb}$ is the bound target electron's momentum. For a ground state hydrogen atom, the bound electron's momentum distribution is sharply peaked near zero. Therefore, if the projectile momentum transfer is large, it will be the dominant term that determines the ejected electron's momentum. In this case, the ADDCSs show a sharp forward peak. If, however, the projectile momentum transfer is small, then the target electron's momentum distribution can significantly influence the ejected electron's momentum. In this case, the ADDCSs exhibit a broad, more uniform peak. In general, a small projectile momentum transfer implies that there is very little change in the projectile's energy and direction, which typically occurs with grazing collisions. In contrast, close collisions between the projectile and target electron often result in large momentum transfer. Therefore, for non-vortex projectiles, the shape of the ADDCS indicates whether ionization occurs though close or grazing collisions.

Figure 1 shows ADDCSs calculated using a Coulomb wave for the ejected electron and neglecting the exchange amplitude. Detailed reasoning for this model choice is described in Appendices A and B. The ADDCSs are shown as a function of ionized electron polar angle for ejection into the x-z plane (scattering plane; $\varphi_e = \pi$) for non-vortex ($\alpha = 0$) and vortex ($\alpha > 0$) projectiles. For non-vortex projectiles, a clear trend is present with increasing ejected electron energy. At low ejected electron energy, the ADDCSs show a broad peak that results from grazing collisions. As energy increases, the peak in the ADDCS narrows due to the dominance of close collisions. For vortex projectiles with small opening angles ($\alpha = 100$ mrad), the ADDCSs are quite similar to those for a non-vortex projectile. This implies that the vortex projectile's uncertainty and non-zero transverse momentum are unimportant. However, for larger opening angles ($\alpha \geq 500$ mrad), the shape of the vortex ADDCSs are very different than the non-vortex ADDCSs, particularly for large ejected electron energies. As the opening angle increases, the peak in the ADDCS broadens and splits into two peaks, indicating that the projectile's uncertainty alters the shape of the ADDCS. Interestingly, the minimum between the two peaks in the ADDCS occurs at the location of the non-vortex ADDCS peak. For example, for an ejected electron energy of 493.2 eV, the non-vortex ADDCS has a peak at $\theta_e = 45°$, but the vortex ADDCSs show a minimum at $\theta_e = 45°$ for $\alpha = 500$ and 785 mrad.

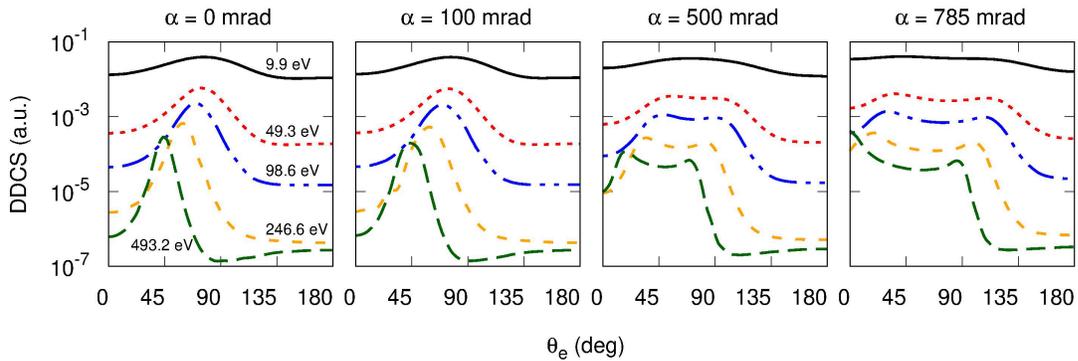

Figure 1 Angular double differential cross sections (ADDCSs) as a function of ejected electron polar angle for non-vortex ($\alpha = 0$) and vortex ($\alpha \neq 0$) projectiles. The incident energy was 1 keV, and ejected electron energies are listed in the figure.

Additional insight into the vortex ADDCS shape can be gained from interpreting the vortex ADDCS as an average over non-vortex ADDCSs for different incident projectile momentum azimuthal angles. In this interpretation, each ADDCS in the average is a result of a projectile with a well-defined incident momentum. Within this interpretation, none of the incident vortex momenta lie along the non-vortex incident momentum direction (+z axis). Therefore, none of the possible vortex momentum transfer vectors are along the direction of the non-vortex momentum transfer direction, which results in the minimum in the vortex ADDCS at the non-vortex ADDCS peak location.

The broadening of the peak in the vortex ADDCS could be due to either the uncertainty in the momentum transfer vector or an enhanced influence of the target electron's momentum, the latter of which will only cause a broadening in the ADDCS if the momentum transfer is small. For $E_2 = 493.2$ eV and $\alpha = 45°$, the vortex momentum transfer vectors are generally much larger than the target electron's momentum ($2 < q < 16$; $k_{eb} < 1$) and therefore it is highly unlikely that the broadening of the ADDCS peak is due to the target electron uncertainty. Additional evidence that the incident vortex projectile's uncertainty causes the broadening in the ADDCS can be found from examination of the component ADDCSs. The shape of these component ADDCSs will indicate whether a projectile with a specific incident momentum causes ionization through a close or grazing collision.

Figure 2 shows the component ADDCSs for all incident projectile azimuthal angles at three different ejected electron energies with the largest opening angle ($\alpha = 45°$). In the heat maps, warmer colors indicate larger cross sections and cooler colors are smaller cross sections. Each row of the heat map is a component ADDCS with well-defined incident projectile momentum azimuthal angle. A few select component ADDCSs are shown to the right of each heat map. Averaging values in each column of the heat map yields the vortex ADDCSs shown in the right panel of Figure 1.

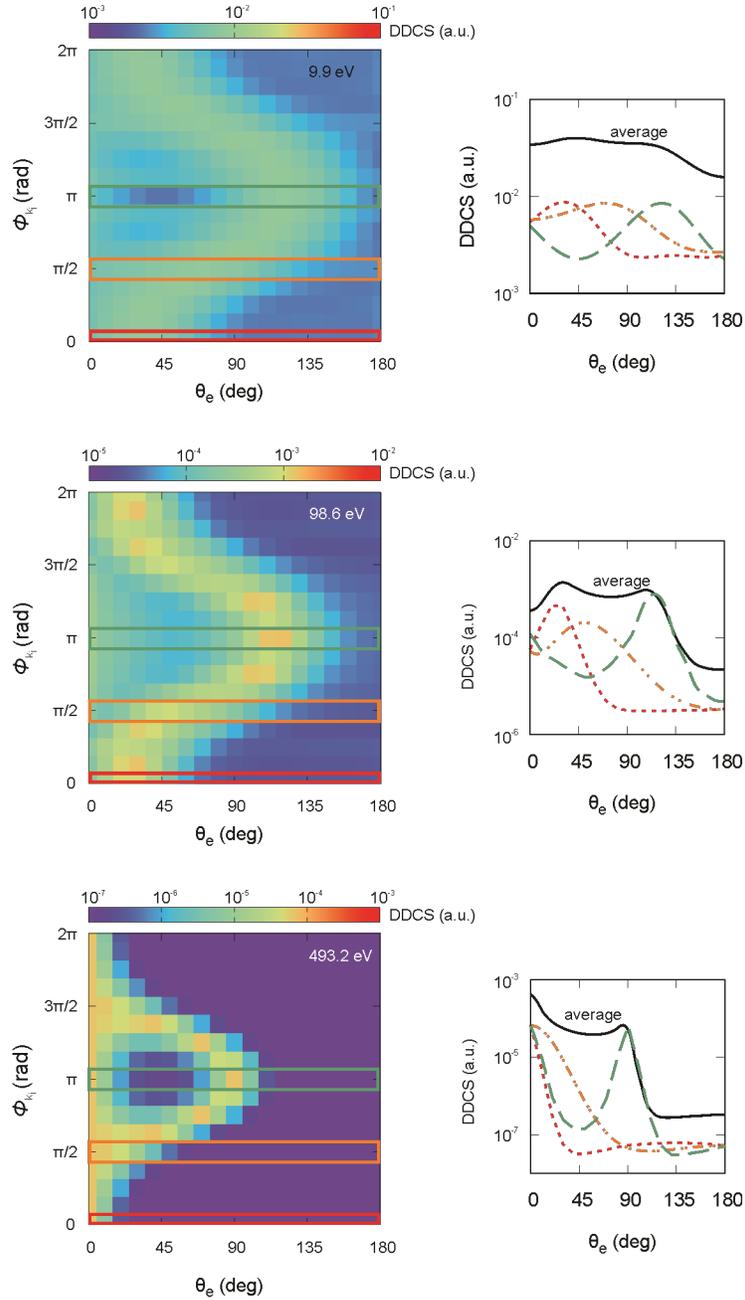

Figure 2 Component angular double differential cross sections (ADDCSs) for ejected electron energies of 9.9 eV (top), 98.6 eV (middle), and 493.2 eV (bottom). The incident energy was 1 keV and the opening angle was $\alpha = 45°$. In the left column, heat maps of the component ADDCSs are plotted as a function of ejected electron angle ($\theta_e$) and incident projectile azimuthal angle ($\phi_{k_i}$). A few selected component ADDCSs are plotted in the right column. The colored boxes highlighting each row in the heat map correspond to the same colored curves in the plots in the right column. Green is $\phi_{k_i} = \pi$, orange is $\phi_{k_i} = \frac{\pi}{2}$, and red is $\phi_{k_i} = 0$. The solid black curve in the right column is the average ADDCS (same as the right panel of Figure 1).

At all energies, the heat maps show that the component ADDCSs are peaked near $\theta_e =$

0° for $\phi_{k_i}$ near 0, while for $\phi_{k_i}$ near $\pi$, they are peaked near $\theta_e = 90°$. As the energy of the ejected electron increases, the peaks in the component ADDCSs become sharper, indicating that the projectile momentum transfer is much larger than the target electron momentum and that ionization occurs via a close collision. At the lowest ejected electron energy of 9.9 eV, the component ADDCSs are broader, which suggests ionization through a grazing collision.

The broad, double peak structure observed in the averaged ADDCSs of Figure 1 for large ejected electron energy and large opening angle can be traced to the individual component ADDCSs shown in Figure 2. Unlike the average, each component ADDCS has a well-defined peak, indicating that the broadening in the average ADDCS is due to the contributions of many different peak locations from each of the component ADDCSs. This is additional evidence that the broadening is caused by the momentum transfer uncertainty, and not an increased influence of the target electron's momentum distribution.

As noted above, the double peak structure for large ejected electron energy and large opening angle results from component ADDCSs with an incident projectile momentum azimuthal angle near $\phi_{k_i} = 0$ and $\phi_{k_i} = \pi$. Although the forward peak in the component ADDCS at $\theta_e = 0$ and $E_2 = 493.2$ eV is sharpest for $\phi_{k_i} = 0$, it is present for all incident momentum azimuthal angles and results in the dominant forward ionization peak observed in the averaged ADDCS (Figure 1). This strong forward ionization peak points to momentum transfer in the forward direction, which results from the projectile being scattered nearly into the perpendicular plane (x-y plane). Thus, the forward ADDCS peak is due to the approximately perpendicular emission of the two final state electrons, with one emitted near the propagation direction (z-axis) and one emitted nearly into the perpendicular plane (x-y plane). The perpendicular emission of two final state electrons with equal energy sharing ($E_2 = 493.2$ eV) is indicative of a binary collision between two equal mass particles with one initially at rest.

The secondary peak at $\theta_e = 90°$ in the component ADDCS is due exclusively to an incident projectile azimuthal angle of $\phi_{k_i} = \pi$. In this case, the incident projectile's momentum lies in the x-z plane, with a transverse component along the -x axis. If it is assumed that the emission is due to a binary collision, then in order for an electron to be ejected at $\theta_e = 90°$, momentum conservation dictates that the projectile must be scattered in the backward (-z) direction. In this case, both final state electrons are again emitted at 90° to each other, with one emitted backward and the other emitted in the perpendicular (x-y) plane.

At low ejected electron energy, the average ADDCS peak is also broadened and shows a slight double peak structure. Examination of the component ADDCSs reveals similar trends to the high ionized electron energy case. In particular, forward scattering results from all incident projectile azimuthal momentum angles, with the largest contribution for $\phi_{k_i} = 0$. The secondary peak is due to the $\phi_{k_i} = \pi$ component ADDCS.

### B. Energy Double Differential Cross Sections

In addition to the angular spectra in Figs. 1 and 2, the energy spectra for ionized electrons also provides insight into the binary encounter process. Under the Bethe conditions, all momentum from the projectile is transferred to the target electron and none is transferred to the nucleus [3]. In this case, the non-vortex EDDCSs exhibit a binary encounter peak at

$$E_{BE}^{NV} = E_i \cos^2 \theta_e - I_p, \tag{18}$$

where $I_p$ is the atomic ionization potential. For helium and H$_2$ targets, it has been shown that the FBA reasonably predicts the binary encounter peak [3].

Figure 3 shows the EDDCSs for vortex and non-vortex projectiles at several different ionized electron angles ($\theta_e = 10, 30, 50, 70°$), which are near the peak of the non-vortex ADDCSs. In this region, the binary encounter approximation is expected to be valid [1], and the FBA should reasonably predict the binary encounter peak. The EDDCSs in Fig. 3 have been calculated using the FBA with plane waves for all free particles, including both the direct and exchange amplitudes. Appendices A and B shows that for EDDCSs, the plane wave approximation is sufficient, but that exchange is crucial at the larger ejected electron energies.

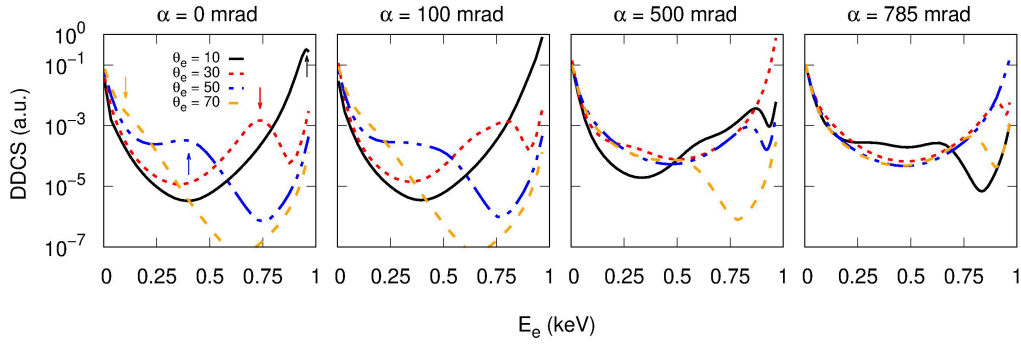

Figure 3 Energy double differential cross sections (EDDCSs) as a function of ejected electron energy for non-vortex ($\alpha = 0$) and vortex ($\alpha \neq 0$) projectiles. The incident energy was 1 keV, and ejected electron energies are listed in the figure. The arrows in the left panel indicate the predicted binary encounter peak location, as given by equation (18).

For non-vortex projectiles, the peak in the EDDCS is located at energies predicted by Eq. (18), which shows that the peak location moves to smaller energies as ionized electron angle increases. The binary encounter peak is most distinct at $\theta_e = 30°$ and $50°$, while for $\theta_e = 10°$ and $70°$, it is less noticeable. At $\theta_e = 10°$, the large cross section near 950 eV is the binary encounter peak. For $\theta_e = 70°$, the slight shoulder in the cross section near 100 eV is the binary encounter peak. In this case, the binary encounter electrons emitted at low energy overlap with those emitted due to soft collisions, washing out any strong peak structure.

For vortex electrons with small opening angle ($\alpha = 100$ mrad), the EDDCSs are similar to the non-vortex EDDCSs. The only noticeable difference is a slight flattening of the binary encounter peaks. As the opening angle increases, the shape of the EDDCS changes, particularly at large ($E_e > 100\ eV$) ejected electron energy. A clear binary encounter peak is not easily identifiable for projectiles with large opening angles, but the peak structures that are present have shifted to larger energy. For $\theta_e = 30°$ and $50°$, an increase in magnitude is observed near $E_e = 1$ keV for $\alpha = 500$ mrad and $\alpha = 785$ mrad, respectively. In contrast, for $\theta_e = 10°$, the magnitude of the EDDCS for large $\alpha$ near 1 keV is much smaller than non-vortex projectiles.

Identification of a clear binary encounter peak in the EDDCSs is difficult due to the average of the vortex cross section over the incident projectile azimuthal angle. However, as was the case with the ADDCSs, insight into the EDDCSs can be found from examining the component EDDCSs. Additionally, Eq. (18) can be generalized for a vortex projectile with a specific $\phi_{k_i}$, leading to a predicted peak at

$$E_{BE}^V = E_i\left[\sin^2\alpha \sin^2\theta_e \cos^2\phi_{k_i} + \cos^2\alpha \cos^2\theta_e - 2\sin\alpha \sin\theta_e \cos\phi_{k_i} \cos\alpha \cos\theta_e\right] - I_p$$

(19)

Figure 4 shows the component EDDCSs for all incident projectile azimuthal angles. As was the case for the component ADDCSs in Fig. 2, warmer colors represent larger cross sections and each row is a component EDDCS with well-defined incident projectile azimuthal angle. Averaging each column of the heatmaps yields the EDDCSs of the right panel in Fig. 3. A few select component EDDCSs ($\phi_{k_i} = 0, \frac{\pi}{2}, \pi$) are shown to the right of each heatmap. The white line superimposed on the heatmaps is the binary encounter peak location as predicted by Eq. (19). This shows that a binary encounter collision leads to the EDDCS peak at large ionized electron energies. In particular, the component EDDCS with $\phi_{k_i} = \pi$ is responsible for the enhancement of the vortex EDDCS near 1 keV. At low ionized electron energies, all incident projectile azimuthal angles contribute approximately equally.

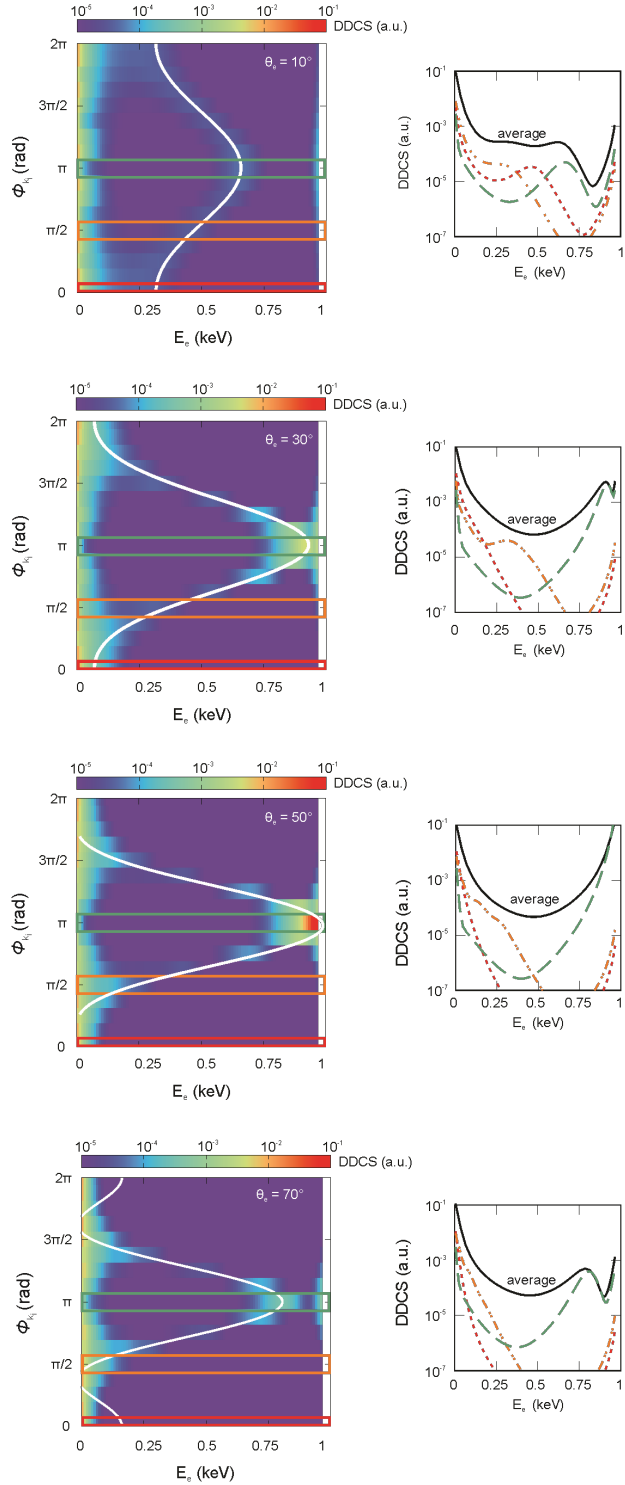

Figure 4 Component energy double differential cross sections (EDDCSs) for ejected electron angles of $\theta_e = 10°$ (top), $\theta_e = 30°$ (second), and $\theta_e = 50°$ (third), and $\theta_e = 70°$ (bottom). The incident energy was 1 keV and the opening angle was $\alpha = 45°$. In the left column, heat maps of the component EDDCSs are plotted as a function of ejected electron energy ($E_e$) and incident projectile azimuthal angle ($\phi_{k_i}$). The solid white line on the heat maps is the binary encounter

peak location as predicted by Eq. (19).  A few selected component EDDCSs are plotted in the right column.  The colored boxes highlighting each row in the heat map correspond to the same colored curves in the plots in the right column.  Green is $\phi_{k_i} = \pi$, orange is $\phi_{k_i} = \frac{\pi}{2}$, and red is $\phi_{k_i} = 0$.  The solid black curve in the right column is the average EDDCS (same as the right panel of Figure 3).

At a small ejected electron angle ($\theta_e = 10°$), the binary encounter peaks at each $\phi_{k_i}$ are less prominent than the low energy peak, indicating that the binary encounter process is less important for electrons ejected in the forward direction.  As ionized electron angle increases, the binary encounter peak for $\phi_{k_i} = \pi$ becomes more prominent and is most noticeable for $\theta_e = 30°$ and 50° near $E_e = 1\ keV$.  This is a direct result of the projectile's transverse momentum, which leads to a momentum transfer direction near 45° and thus an increased prevalence of binary encounter electrons for these kinematics.

## 4. Discussion and Conclusion

We have presented theoretical angular and energy double differential cross section spectra for collisions between electron vortex projectiles and hydrogen atoms.  Our results show that for small opening angles, the angular and energy DDCSs are similar for vortex and non-vortex projectiles.  However, for large opening angles, the vortex DDCSs show noticeable differences compared to the non-vortex DDCSs.

The sharpness of the peak in the non-vortex ADDCS indicates whether ionization occurs through grazing (broad peak) or close collisions (sharp peak).  In the vortex ADDCS, the peak broadens and at large vortex opening angles becomes a double peak.  However, this broadening is not an indication of increased importance of grazing collisions.  Rather, it is a result of the incident projectile's momentum uncertainty and non-zero azimuthal momentum.  In general, the changes in shape of the vortex ADDCS partially obscure the evidence for ionization mechanisms, although analysis of the component ADDCSs serves to clarify the mechanisms.

The component ADDCSs predict that, as was the case for non-vortex projectiles, for large ejected electron energy, ionization occurs through close collisions. In contrast, for low ejected electron energy, ionization occurs through grazing collisions.

For non-vortex projectiles, when the Bethe condition is satisfied, the EDDCSs show a distinct binary encounter peak. For vortex projectiles, this binary encounter peak is not readily observable and is again obscured due to the incident projectile's momentum uncertainty. The magnitude of the vortex EDDCSs are increased at large ejected electron energy and angle compared to their non-vortex counterparts. This enhancement is due to the vortex projectile's transverse momentum and can be traced to projectiles with azimuthal momentum angle $\phi_{k_i} = \pi$.

The results presented here are theoretical predictions for ejected electron angular and energy spectra. They present an opportunity to identify specific ionization mechanisms and to test the validity of a binary collision model for vortex projectiles. In particular, the predictions of our model could be tested by comparison with experiment. For example, the double peak structure in our ADDCSs at large opening angles and ionized electron energy is a direct result of ionization through close collisions. If the ionization mechanism is different, then experimental ADDCSs will not show this double peak structure. Additionally, the validity of the model for vortex projectiles can be tested by comparison with experimental EDDCSs, whose magnitude at large ejected electron energies will indicate whether or not binary encounter collisions are important for these kinematics. Unfortunately, there is currently no experimental data available for vortex collisions with atomic targets. However, our results indicate that measurement of angular and energy spectra can provide insight into electron vortex ionization mechanisms, which may help to advance the applications of sculpted electrons. We hope that this study provides motivation for future experiments.

# Appendix A

Proper calculation of the DDCS within the FBA includes both the direct and exchange transition amplitudes as shown in Eq. (3). However, calculation of the exchange amplitude is numerically intensive if the ejected electron wave function is a Coulomb wave. For highly asymmetric energy sharing with a slow ejected electron, the exchange amplitude can be safely neglected, but this may not be the case for fast ejected electrons. Therefore, it is important to establish under what conditions the exchange amplitude needs to be included.

Figures A1 and A2 show the effect of the exchange amplitude on the ADDCSs and EDDCSs respectively. The calculations were performed within the FBA model with the ejected electron treated as a plane wave. These results show that exchange can safely be neglected for the ADDCSs, even at ejected electron energies as high as 493.2 eV. However, exchange is important in the EDDCSs at large ejected electron energies above 500 eV. These results informed our decision to neglect exchange in the ADDCSs, but include it in the EDDCSs.

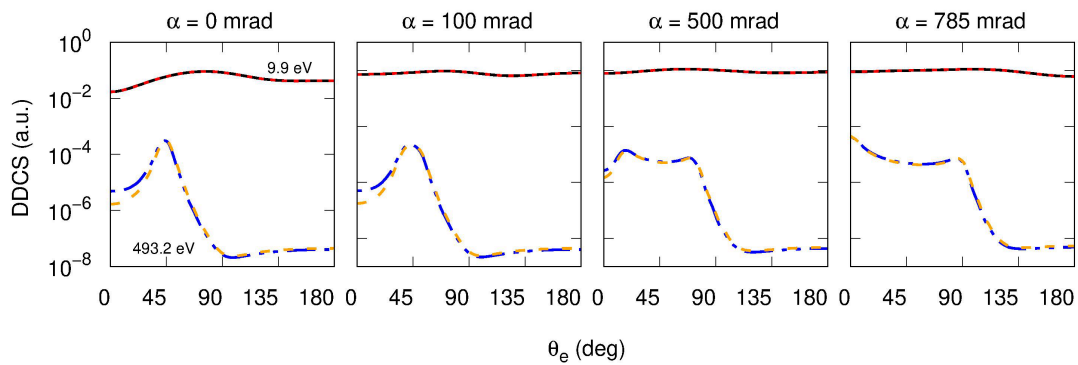

Figure A1 Angular double differential cross sections (ADDCSs) using the FBA model with a plane wave treatment of the ejected electron. The solid black and dash-double dotted blue lines are calculations including both the direct and exchange amplitudes. The dotted red and dashed yellow lines are calculations including only the direct amplitude. The incident energy was 1 keV and the ejected electron energies are shown in the figure.

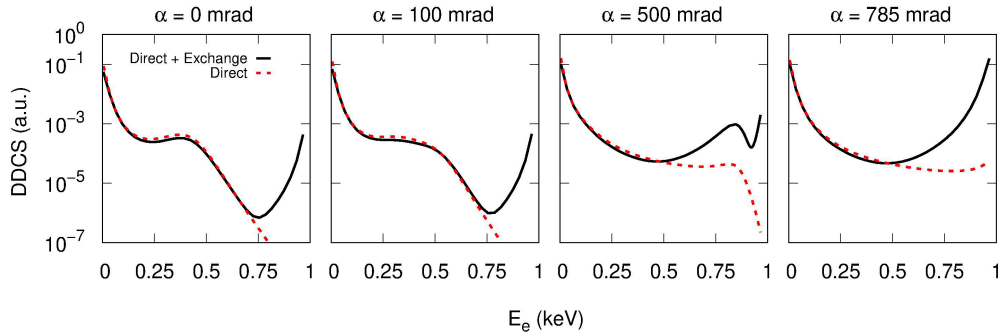

Figure A2 Energy double differential cross sections (EDDCSs) using the FBA model with a plane wave treatment of the ejected electron. The solid black lines are calculations including both the direct and exchange amplitudes. The dotted red lines are calculations including only the direct amplitude. The incident energy was 1 keV and the ionized electron angle was $\theta_e = 50°$.

## Appendix B

Within the FBA model, the ejected electron can be modeled as either a plane wave (Eq. (7)) or a Coulomb wave (Eq. (8)). To test the importance of the Coulomb distortion on the ejected electron, we calculated both ADDCSs and EDDCSs for plane wave and Coulomb wave treatments. Figures B1 and B2 show the effect of the Coulomb distortion on the ADDCSs and EDDCSs respectively. The calculations were performed within the FBA model neglecting exchange. These results show that a Coulomb wave treatment is important in the ADDCSs, particularly at low ejected electron energies. However, a Coulomb wave treatment of the ejected electron is generally unimportant in the EDDCSs. These results informed our decision to treat the ejected electron as a Coulomb wave in the ADDCSs, but as a plane wave in the EDDCSs.

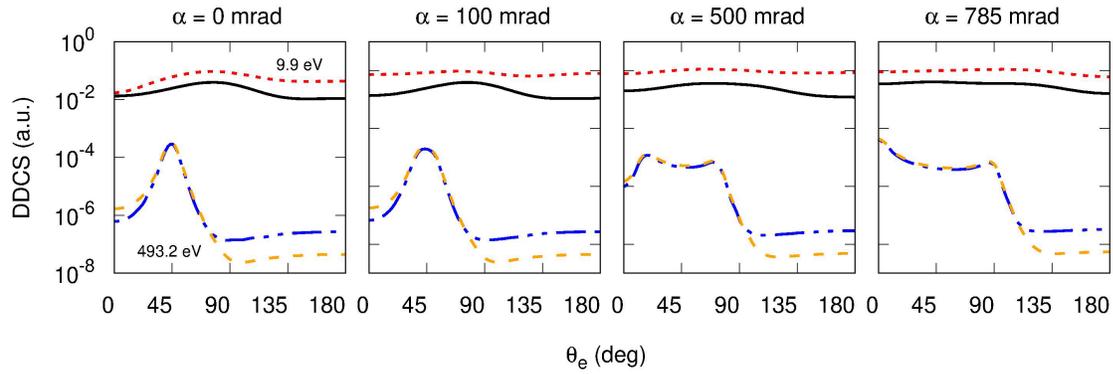

Figure B1 Angular double differential cross sections (ADDCSs) using the FBA model with either a plane wave or Coulomb wave treatment of the ejected electron. The exchange amplitude was not included. The solid black and dash-double dotted blue lines are calculations using a Coulomb wave. The dotted red and dashed yellow lines are calculations using a plane wave. The incident energy was 1 keV and the ejected electron energies are shown in the figure.

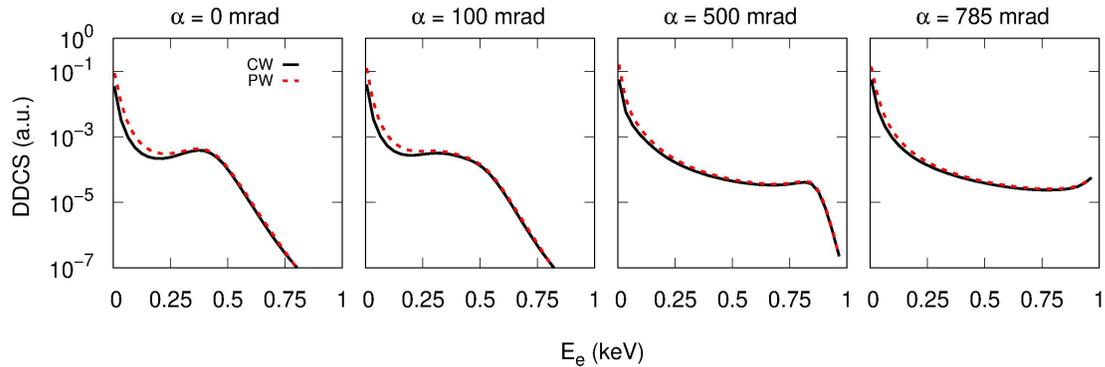

Figure B2 Energy double differential cross sections (EDDCSs) using the FBA model with either a plane wave or Coulomb wave treatment of the ejected electron. The exchange amplitude was neglected. The solid black lines are calculations using a Coulomb wave. The dotted red lines are calculations using a plane wave. The incident energy was 1 keV and the ionized electron angle was $\theta_e = 50°$.

## Acknowledgements

We gratefully acknowledge the support of the National Science Foundation under Grant No. PHY-1912093.